\documentclass[letter]{spie}
\usepackage{amsmath,amsfonts,amssymb}
\usepackage{graphicx}
\usepackage[colorlinks=true, allcolors=blue]{hyperref}
\usepackage[square, numbers]{natbib}
\usepackage{enumitem}
\setlist[itemize]{noitemsep}
\usepackage{multirow}
\usepackage{microtype}
\usepackage{mathptmx} 
\usepackage{caption}
\usepackage{booktabs}
\usepackage{pgfplots}
\usepackage{pgfplotstable}
\usepgfplotslibrary{polar}
\pgfplotsset{compat=1.18}

\title{LALE: Lightweight-Transformer Architecture for Land-Cover Estimation}
\author{Ümit Mert Çağlar}
\author{Alptekin Temizel}
\affil{Graduate School of Informatics, METU, Ankara, Turkey}

\authorinfo{Further author information: (Send correspondence to Ümit Mert Çağlar)\\Ümit Mert Çağlar: E-mail: mecaglar@metu.edu.tr\\Alptekin Temizel: E-mail: atemizel@metu.edu.tr}

\begin{document} 
\maketitle

\begin{abstract}
% Complete-revised
Semantic segmentation of remote sensing imagery requires models that capture both global context and local detail under tight computational budgets. Prior work typically optimizes for one of these axes: attention for global context, convolution for local detail, or compactness for efficiency. While hybrid approaches aim to capture both, they require architectural changes and encoder backbones with computational overhead, limiting efficiency and performance. We present LALE (Lightweight-transformer Architecture for Land-cover Estimation), an end-to-end remote sensing image segmentation architecture, that bifurcates its encoder by resolution: lightweight ConvMixer stages handle high-resolution local features, while transformer stages handle low-resolution global context,  confining the quadratic cost of self-attention to deep, downsampled feature maps. An all-MLP multi-scale decoder, together with RMSNorm and StarReLU throughout, further reduces compute and parameter count. On the large-scale ARAS400k remote-sensing segmentation benchmark, LALE establishes a strong efficiency-performance trade-off against CNN, transformer, and hybrid baselines. Our smallest variant, (just 1.6M parameters), reaches within 2.6 F1 points of the best baseline (UPerNet) while using $4.5\times$ fewer parameters, $7\times$ less storage, $17\times$ fewer GMACs, and delivering $1.8\times$ higher throughput. The codebase for LALE is publicly available at \href{https://github.com/caglarmert/LALE}{https://github.com/caglarmert/LALE}.

\end{abstract}


\section{Introduction}
\label{sec:intro}
% Complete-revised
Semantic segmentation of remote sensing images is crucial for a wide range of applications, including environmental monitoring, urban planning, disaster awareness and precise land-cover estimation. Convolutional Neural Networks (CNNs) bring strong inductive biases and computational efficiency, but their limited receptive fields restrict long-range reasoning. Vision transformers model global context effectively through self-attention, yet their quadratic cost in spatial resolution makes them expensive. This is a particularly important drawback in remote sensing, where inputs are high-resolution, throughput requirements are large, and deployment platforms are often resource-constrained. Hybrid architectures aim to combine the strengths of both families, but most are assembled from generic ImageNet backbones paired with parameter-heavy decoders, inheriting overhead that is not justified by the demands of the segmentation task itself. As a result, the efficiency-performance trade-off in remote sensing segmentation remains an open problem. 

\begin{figure}
    \centering
    \includegraphics[trim={0 570 0 000},clip,width=1\linewidth]{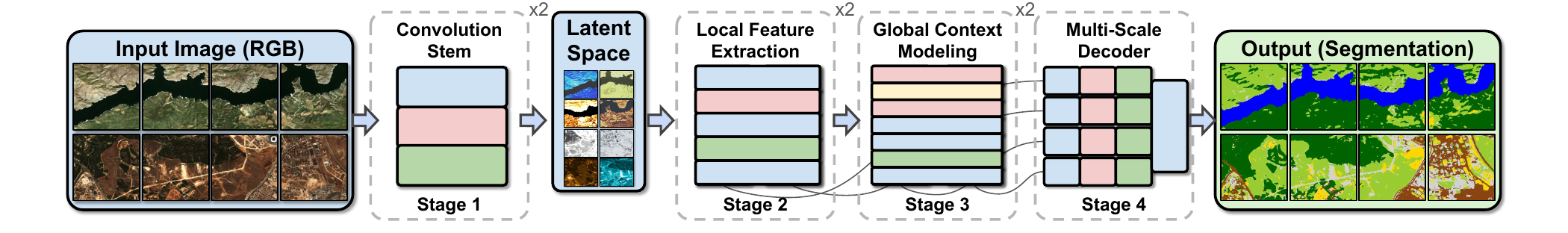}
    \caption{Overall approach for the Lightweight-Transformer Architecture for Land-cover Estimation (LALE)}
    \vspace{-10pt}
    \label{fig:overall}
\end{figure}

In this work, we introduce Lightweight-Transformer Architecture for Land-Cover Estimation (LALE), an end-to-end segmentation model designed specifically for this trade-off, illustrated in Figure~\ref{fig:overall}. LALE bifurcates its encoder by resolution: lightweight ConvMixer stages process high-resolution feature maps where local detail dominates, while transformer stages operate on deeper, downsampled feature maps where global context matters and the quadratic cost of self-attention is tractable. A lightweight all-MLP multi-scale decoder replaces conventional heavy upsampling heads, and operation-level choices (RMSNorm in place of LayerNorm, StarReLU in place of GELU, and $3\times3$ stride-2 convolutions in the stem) further reduce compute without degrading representational capacity.

Extensive experiments on the large-scale ARAS400k benchmark show that LALE achieves a strong performance-efficiency trade-off. Against CNN baselines, LALE matches segmentation accuracy at a fraction of the parameter and compute budget; against dense-prediction transformer baselines, it delivers comparable accuracy with one to two orders of magnitude lower compute and substantially higher throughput, making it well-suited to real-time and resource-constrained deployment.

\newpage
\noindent Our contributions are:\vspace{-10pt}
\begin{itemize}
    \item We introduce LALE, a lightweight attention-convolution hybrid hierarchical architecture for semantic segmentation of remote sensing images, which optimize the computational efficiency and segmentation task performance.
    \item LALE, a resolution-bifurcated convolution-transformer hybrid encoder paired with an all-MLP multi-scale decoder, designed end-to-end for remote sensing semantic segmentation.
    \item A set of operation-level efficiency choices (RMSNorm, StarReLU, and small-kernel strided downsampling) that together reduce compute and memory footprint without sacrificing accuracy. 
    \item A comprehensive benchmark on ARAS400k spanning CNN, transformer, and hybrid architectures, establishing an extended reference point for future work on this dataset.
    \item Empirical evidence that LALE occupies a favorable region of the accuracy-efficiency Pareto frontier.
\end{itemize}
\vspace{-20pt}

\section{Related Work}
%Complete-Revised
Semantic segmentation in remote sensing presents unique challenges, including extreme scale variations, high-resolution processing demands, and complex, heterogeneous backgrounds. To address these, early architectural paradigms relied heavily on CNNs, with UNet~\cite{ronneberger2015u} establishing a robust baseline through a symmetrical encoder-decoder structure and lateral skip connections to recover spatial hierarchies. Expanding on this foundation to better handle dynamic multi-scale fusion, UNet++~\cite{zhou2018unet++} introduced densely connected, cross-fused skip connections spanning varying scaling ratios, enabling the hyper-fine segmentation of complex objects. Parallel structural adaptations, such as Linknet~\cite{chaurasia2017linknet}, utilized lightweight encoders and additive skip connections to streamline memory utilization and accelerate spatial processing. The Feature Pyramid Network (FPN)~\cite{lin2017feature} emerged to inject high-level abstract semantics across multiple spatial resolutions via a bidirectional structural pathway. Pyramid Attention Network (PAN)~\cite{li2018pyramid} merged CNNs with spatial attention mechanisms to fuse multi-scale context and provide precise pixel-level attention for small targets without relying on dilated convolutions. 

Concurrently, the DeepLabV3~\cite{chen2017rethinking} series advanced multi-scale context capture using Atrous Spatial Pyramid Pooling (ASPP), with DeepLabV3+~\cite{chen2018encoder} integrating a dedicated decoder and depthwise separable convolutions to recover remarkably sharp architectural boundaries. The Unified Perceptual Parsing Network (UPerNet)~\cite{xiao2018unified} coupled an FPN with a Pyramid Pooling Module (PPM) to learn deep global pairwise relationships, actively mapping intra-class consistencies across expansive land-cover zones. This continuous drive toward broader spatial understanding ultimately set the stage for attention-driven architectures like SegFormer~\cite{xie2021segformer}, which adopted a positional-encoding-free Mix Transformer (MiT) encoder and a lightweight MLP decoder, granting the model an effectively infinite receptive field to capture long-range environmental dependencies.

Recent advancements in vision transformers (ViTs) have focused on enhancing the training stability and scalability of base architectures. DeiT3~\cite{touvron2022deit} demonstrates that by refining training recipes, incorporating advanced data augmentation and regularization, standard ViTs can significantly outperform previous iterations without increasing model complexity. Similarly, SwinV2~\cite{liu2022swin} introduces a post-normalization configuration and log-spaced relative position biases, achieving better stability and effectively adapts to large-scale visual recognition tasks. To bridge the gap between local and global context, several models utilize hybrid attention mechanisms such as MaxViT~\cite{tu2022maxvit}, introducing Multi-Axis Attention that decomposes the standard attention mechanism into local window attention and global sparse grid attention. This allows for global-local spatial interactions with linear complexity relative to the image size, making it highly versatile across different scales.

A significant trend in transformer research is the optimization for real-time inference on edge devices. EfficientFormer~\cite{li2023rethinking} re-evaluates the design bottlenecks of ViTs, proposing a dimension-consistent design that integrates the speed of Convolutional Neural Networks (CNNs) with the performance of transformers. FastViT~\cite{vasu2023fastvit} builds upon this by utilizing structural reparameterization and a novel RepMixer block. It minimizes latency by reducing memory access costs, making it one of the most efficient hybrid models for mobile deployment.

\section{Method}
\label{sec:method}
\begin{figure}[ht]
    \centering
    \includegraphics[trim={0 110 20 0},clip,width=0.8\linewidth]{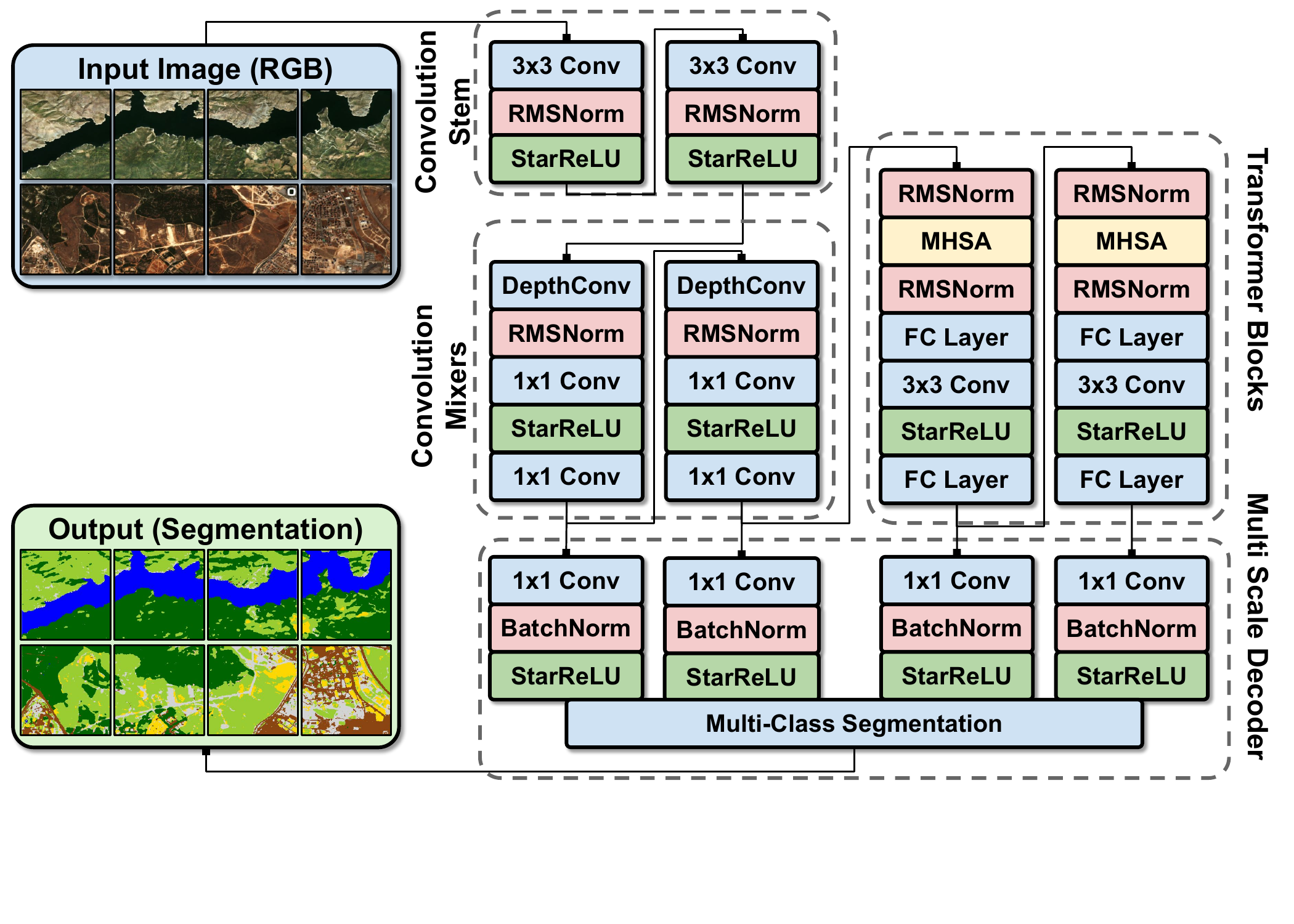}
    \caption{Architecture of the Lightweight-Transformer Architecture for Land-cover Estimation (LALE) consists of four sections, convolution stem, convolution mixers acting as local feature extractor, Transformer blocks for global context modeling and multi-scale segmentation decoder. Convolution stem and mixers utilize efficient convolution approaches while transformer blocks utilize computation-intensive multi-head self-attention (MHSA) and multi-scale decoder fully connected dense layers. Multi-scale decoder extract information from local and global features representing fine and coarse details.}
    \vspace{-10pt}
    \label{fig:detailed_lale}
\end{figure}

% Complete-revised
We introduce a lightweight attention-convolution hybrid hierarchical architecture (LALE) for remote sensing semantic segmentation (Figure~\ref{fig:detailed_lale}). LALE integrates global context modeling capabilities of attention-based architectures (transformers) and the local inductive biases of CNNs. The architecture contains three distinct modules, an overlapping patch embedding stem, a four-stage bifurcated hybrid encoder and a lightweight multi-scale decoder. %LALE is an end-to-end model for semantic segmentation tasks with true color RGB images as its input and color-encoded semantic segmentation maps as its output. We tweak LALE for the native resolution of ARAS400k dataset, $256 \times 256$, and it is possible to change the overall architecture with the model configuration.

\subsection{Overall Architecture}

Our architecture begins with a given input image $\mathbf{I} \in \mathbb{R}^{3 \times H \times W}$ with $H=W=256$, the network first extracts dense early-stage features with its convolutional stem. The stem consists of two sequential $3 \times 3$ convolutions each with a stride of $2$ and a padding of $1$, where each convolution is followed by RMSNorm and a StarReLU activation. Because each convolution halves the spatial resolution, the stem reduces the input by a factor of four overall. The first convolution projects the three input channels to an intermediate dimension, and the second produces the stem output $\mathbf{F}_0 \in \mathbb{R}^{C_1 \times \frac{H}{4} \times \frac{W}{4}}$ with $C_1 = 32$ is obtained with the convolution, The use of small $3\times3$ kernels with stride 2 provides overlapping receptive fields between adjacent output positions, giving the stem the locality benefits of a patch-embedding layer at substantially lower cost than the $7\times7$ kernels typically used for this purpose.

The feature map $\mathbf{F}_0$ is subsequently processed through by a four-stage hierarchical encoder, generating multi-scale features $\mathbf{F}_i$ for $i \in \{1, 2, 3, 4\}$ at resolutions $\frac{H}{2^{i+1}} \times \frac{W}{2^{i+1}}$ and channel dimensions $C_i \in \{32, 64, 128, 256\}$. First two stages utilize ConvMixer blocks for local feature extraction and last two stages utilize transformer blocks for global context modeling. A $3 \times 3$ convolution with a stride of $2$ followed by RMSNorm is used for spatial downsampling between stages. Local processing stages (stage 1 and 2) operate on highest resolution feature maps, using efficient ConvMixer blocks to extract dense and localized representations. Global processing stages (stage 3 and 4) operate on downsampled feature maps, using the transformer blocks. The $O(N^2)$ self-attention computation bottleneck of transformers is effectively restricted to the low-resolution latent spaces at the last stages.

\subsection{Efficient Normalization and Activation}

We replace "standard" normalizations and activations used in the transformer architectures with highly efficient alternatives to maintain a smaller computational footprint without representation capacity degradation. We utilize RMSNorm~\cite{zhang2019root} to decrease the computational overhead of the standard layer normalization~\cite{ba2016layer}. RMSNorm normalize the activations strictly based on the root mean square of the features, improving efficient by omitting the re-centering operation. We employ StarReLU~\cite{yu2023metaformer} to reduce the activation FLOPs and achieve better performance. StarReLU eliminates the need for expensive exponential operations found in GELU~\cite{hendrycks2016gaussian} or Swish~\cite{ramachandran2017searching} activation functions.

\subsection{Hybrid Encoder Blocks}
\textbf{ConvMixer Block (Early Stages):} In the high-resolution early stages (stage 1 and 2), global self-attention calculation requires much more computation (compared to low-resolution later stages) thus we introduced a ConvMixer block~\cite{trockman2022patches}. For an input tensor $\mathbf{Z}$, the block performs spatial mixing via a depthwise convolution, followed by channel mixing via pointwise convolutions.

\noindent\textbf{Transformer Block (Late Stages):} In the deeper, low-resolution stages (stage 3 and 4), we utilize attention blocks (transformer) to capture long-range semantic dependencies. The block integrates a Multi-Head Attention (MHA) module and a Convolutional Multi-Layer Perceptron (ConvMLP)~\cite{li2023convmlp}. The ConvMLP injects spatial inductive bias into the Transformer by inserting a $3 \times 3$ depthwise convolution between the linear feature expansion and reduction layers.

\subsection{Lightweight Multi-Scale Decoder}

We design an efficient decoder, as proved by many works starting with dense prediction for semantic segmentation with Fully Connected Networks (FCN)~\cite{long2015fully} and more recent works such as Simple and Efficient Design for Semantic Segmentation with Transformers (SegFormer~\cite{xie2021segformer}). Rather than relying on complex, parameter-heavy upsampling blocks, our multi-scale decoder aggregates the multi-scale features $\{\mathbf{F}_1, \mathbf{F}_2, \mathbf{F}_3, \mathbf{F}_4\}$ using simple transformations. First, pointwise ($1 \times 1$) convolutions project each $\mathbf{F}_i$ to a unified channel dimension $C_{\text{dec}} = 128$. Next, features from deeper stages are bilinearly upsampled to the spatial resolution of $\mathbf{F}_1$ (i.e., $\frac{H}{4} \times \frac{W}{4}$). The aligned features are concatenated along the channel axis and fused. To maximize final prediction accuracy, we employ Batch Normalization in the fusion stage instead of RMSNorm since the previous stages ensure high model efficiency. Finally, a dropout layer ($p=0.1$) provides regularization before a final $1 \times 1$ convolution predicts the unnormalized logits for the $N$ target classes. The outputs are bilinearly interpolated back to the original $H \times W$ resolution.

\subsection{Efficient Kernel Infrastructure}
Vision transformers often employ $7 \times 7$ kernels for their patch embeddings as larger kernels enable broader context, while smaller kernels are oriented towards local details. Our convolution-attention hybrid multi-stage encoder and depth-agnostic convolution layers are already competent for global context. Utilizing smaller kernels ($3 \times 3$) with a comparatively large stride ($2$) we remove the redundancy for broader-context capability, improve the computation efficiency for high-resolution front stages and improve local detail capability of our architecture. Kernel size $3 \times 3$ with stride $2$ is $22\times$ more efficient than $7 \times 7$ with stride $1$, $5.4\times$ more efficient than $7 \times 7$ with stride $2$ and $1.4\times$ more efficient than $7 \times 7$ with stride $4$.

\section{Results}
%Complete-Revised

We separate three model groups, first group consist of CNN-based, attention-based and hybrid architectures, the second group is our proposed method LALE with four different sizes and the third group is for dense prediction transformer applications with different vision transformer encoders. We utilized EfficientNet~\cite{tan2019efficientnet} as the encoder for the first group, which is a more efficient, better-performing and newer model than the ResNet~\cite{he2016deep} architecture, reducing the number of parameters and model size by nearly half.

Table~\ref{tab:segmentation_combined} and Figure~\ref{tab:combined_figures_tradeoff} reveals clear trade-offs between the three model groups. CNN-based architectures achieve the highest segmentation performance, with UPerNet and Unet leading in F1-score and IoU, but at the cost of moderate computational complexity. In contrast, the proposed LALE models offer a significantly improved efficiency-performance balance. Despite having substantially fewer parameters and lower GMACs, LALE variants achieve competitive results, with only a small drop in accuracy compared to top CNN models. Moreover, they provide the highest throughput, making them well-suited for real-time and resource-constrained scenarios.

Transformer-based models, while competitive in performance, incur a disproportionately high computational cost. Their large model sizes, high GMACs, and low throughput result in poor efficiency scaling, with only marginal accuracy gains over lighter architectures. Overall, LALE achieves the most favorable trade-off, maintaining competitive performance while drastically reducing computational requirements.

\begin{table}[ht]
\small
\renewcommand{\arraystretch}{0.9}
\centering
\caption{Comprehensive performance and efficiency comparison of segmentation architectures. Performance values in \%.}
\label{tab:segmentation_combined}
\begin{tabular}{lccccc|cccc}
\toprule
\textbf{Architecture} 
& \textbf{F1} 
& \textbf{Accuracy} 
& \textbf{Precision} 
& \textbf{Recall} 
& \textbf{IoU} 
& \textbf{Size} 
& \textbf{Throughput} 
& \textbf{Params} 
& \textbf{GMACs} \\
&  
& 
&  
& 
& 
& (MB)
& \textbf{(sample/sec)} 
& (M)
& \\
\midrule
DeepLabV3      & 75.23 & 84.22 & 76.00 & 74.58 & 63.01 & 28.04 & 105,217 & 7.3  & 6.44 \\
DeepLabV3+     & 76.37 & 84.94 & 77.34 & 75.48 & 64.30 & 18.90 & 90,321  & 4.9  & 1.46 \\
FPN            & 76.38 & 85.00 & 76.68 & 76.14 & 64.35 & 22.13 & 97,817  & 5.8  & 2.51 \\
Linknet        & 75.45 & 84.09 & 76.21 & 74.76 & 63.21 & 16.06 & 97,157  & 4.2  & 0.58 \\
PAN            & 76.12 & 84.57 & 76.18 & 76.14 & 63.93 & 15.80 & 83,675  & 4.1  & 0.98 \\
Unet           & 77.23 & 85.09 & 77.13 & 77.53 & 65.28 & 24.02 & 80,002  & 6.3  & 3.05 \\
UnetPlusPlus   & 76.86 & 85.00 & 77.17 & 76.60 & 64.90 & 25.24 & 85,047  & 6.6  & 5.62 \\
UPerNet        & 77.31 & 85.53 & 77.83 & 76.84 & 65.42 & 44.49 & 89,889  & 11.6 & 13.62 \\
Segformer      & 76.47 & 84.82 & 76.05 & 77.10 & 64.44 & 17.23 & 85,491  & 4.5  & 2.05 \\
\midrule
LALE-S1  & 74.69 & 83.16 & 75.18 & 74.26 & 62.25 & 5.98  & 160,253 & 1.6 & 0.59 \\
LALE-S2  & 75.88 & 84.12 & 76.39 & 75.42 & 63.67 & 9.97  & 117,105 & 2.6 & 0.78 \\
% LALE-S3  & 75.83 & 84.17 & 76.41 & 75.30 & 63.63 & 13.95 & 104,228 & 3.7 & 0.97 \\
% LALE-S4  & 75.54 & 83.82 & 76.20 & 74.93 & 63.27 & 17.94 & 86,537  & 4.7 & 1.16 \\
\midrule
EffFormer-L1 & 74.24 & 83.13 & 73.33 & 75.64 & 61.81 & 113.97 & 2,620 & 29.8  & 23.17 \\
EffFormer-L3 & 75.23 & 83.96 & 73.83 & 77.17 & 62.97 & 187.75 & 2,060 & 49.1  & 25.85 \\
EffFormer-L7 & 75.35 & 84.00 & 74.36 & 76.89 & 63.13 & 383.22 & 2,691 & 100.3 & 32.16 \\
DeiT3-Base         & 76.10 & 84.53 & 75.63 & 76.74 & 63.97 & 446.64 & 6,156 & 117.1 & 39.89 \\
MaxViT-Tiny        & 75.82 & 84.46 & 75.55 & 76.24 & 63.64 & 232.26 & 1,589 & 60.8  & 33.13 \\
FastViT-SA12       & 74.71 & 83.36 & 74.29 & 75.32 & 62.32 & 111.71 & 3,980 & 29.2  & 23.40 \\
FastViT-MCI0       & 75.53 & 83.96 & 74.53 & 76.96 & 63.33 & 111.11 & 2,714 & 29.1  & 23.75 \\
\bottomrule
\end{tabular}
\end{table}

\begin{table}[ht]
    \centering
    \begin{tabular}{cc}
        %\begin{figure}[ht]
%\centering
\begin{tikzpicture}
\begin{axis}[
    width=8cm,
    height=6cm,
    xlabel={Parameters (Millions)},
    ylabel={F1-score (\%)},
    grid=major,
    xmin=0, xmax=13,
    ymin=74, ymax=78,
    legend pos=south east,
    legend style={nodes={scale=0.9, transform shape}}
]

% Baseline models
\addplot[
    only marks,
    mark=*,
    color=blue,
    mark size=2.5pt
] coordinates {
    (7.31, 75.23)  % DeepLabV3
    (4.91, 76.37)  % DeepLabV3Plus
    (5.76, 76.38)  % FPN
    (4.17, 75.45)  % Linknet
    (4.10, 76.12)  % PAN
    (6.25, 77.23)  % Unet
    (6.57, 76.86)  % UnetPlusPlus
    (11.61, 77.31) % UPerNet
    (4.47, 76.47)  % Segformer
};
\addlegendentry{CNN-based Models}
% Proposed Models (LALE variants)
\addplot[
    only marks,
    mark=triangle*,
    color=red,
    mark size=4pt
] coordinates {
    (1.57, 74.69) % LALE-S1-PT
    (2.61, 75.88) % LALE-S2-PT
    % (3.66, 75.83) % LALE-S3-PT
    % (4.70, 75.54) % LALE-S4-PT
};
\addlegendentry{LALE Models}

% --- Node Labels ---
% Adjusting anchors (south, north, east, west) to prevent text overlap

\node[anchor=south] at (axis cs:7.31, 75.23) {\scriptsize DeepLabV3};
\node[anchor=north west] at (axis cs:4.91, 76.37) {\scriptsize DeepLabV3+};
\node[anchor=south] at (axis cs:5.76, 76.38) {\scriptsize FPN};
\node[anchor=north east] at (axis cs:4.17, 75.45) {\scriptsize Linknet};
\node[anchor=north west] at (axis cs:4.10, 76.12) {\scriptsize PAN};
\node[anchor=south east] at (axis cs:6.25, 77.23) {\scriptsize Unet};
\node[anchor=north west] at (axis cs:6.57, 76.86) {\scriptsize Unet++};
\node[anchor=south] at (axis cs:11.61, 77.31) {\scriptsize UPerNet};
\node[anchor=east] at (axis cs:4.47, 76.47) {\scriptsize Segformer};

% Labels for Proposed Models
\node[anchor=north, text=red] at (axis cs:1.57, 74.69) {\scriptsize S1};
\node[anchor=north, text=red] at (axis cs:2.61, 75.88) {\scriptsize S2};
% \node[anchor=north, text=red] at (axis cs:3.66, 75.83) {\scriptsize S3};
% \node[anchor=north, text=red] at (axis cs:4.70, 75.54) {\scriptsize S4};

\end{axis}
\end{tikzpicture}
%\caption{Trade-off between F1-score performance and architectural complexity (Parameters in millions). Our smallest proposed model pushes the Pareto frontier in the highly resource-constrained regime.}
%\label{fig:f1_vs_params}
%\end{figure} & 
%\begin{figure}[ht]
\centering
\begin{tikzpicture}
\begin{axis}[
    width=8cm,
    height=6cm,
    xlabel={Parameters (Millions)},
    ylabel={F1-score (\%)},
    grid=major,
    xmode=log,
    log basis x={10},
    xmin=1, xmax=130,
    ymin=74, ymax=77,
    legend pos=north west,
    legend style={nodes={scale=0.9, transform shape}}
]

% Baseline models (NEW ENCODERS)
\addplot[
    only marks,
    mark=*,
    color=blue,
    mark size=2.5pt
] coordinates {
    (49.13, 75.23)  % EfficientFormer-L3
    (29.83, 74.24)  % EfficientFormer-L1
    (117.08, 76.10) % DeiT3-Base
    (60.80, 75.82)  % MaxViT-Tiny
    (100.32, 75.35) % EfficientFormer-L7
    (29.24, 74.71)  % FastViT-SA12
    (29.07, 75.53)  % FastViT-MCI0
};
\addlegendentry{Transformer Models}

% Proposed Models (UNCHANGED)
\addplot[
    only marks,
    mark=triangle*,
    color=red,
    mark size=4pt
] coordinates {
    (1.57, 74.69) % LALE-S1-PT
    (2.61, 75.88) % LALE-S2-PT
    % (3.66, 75.83) % LALE-S3-PT
    % (4.70, 75.54) % LALE-S4-PT
};
\addlegendentry{LALE Models}

% --- Labels (Baseline) ---
\node[anchor=east] at (axis cs:49.13, 75.23) {\scriptsize EffFormer-L3};
\node[anchor=south east] at (axis cs:29.83, 74.24) {\scriptsize EffFormer-L1};
\node[anchor=south east] at (axis cs:117.08, 76.10) {\scriptsize DeiT3-B};
\node[anchor=south] at (axis cs:60.80, 75.82) {\scriptsize MaxViT-T};
\node[anchor=south east] at (axis cs:100.32, 75.35) {\scriptsize EffFormer-L7};
\node[anchor=west] at (axis cs:29.24, 74.71) {\scriptsize FastViT-SA12};
\node[anchor=south east] at (axis cs:29.07, 75.53) {\scriptsize FastViT-MCI0};

% Labels for Proposed Models
\node[anchor=north, text=red] at (axis cs:1.57, 74.69) {\scriptsize S1};
\node[anchor=north, text=red] at (axis cs:2.61, 75.88) {\scriptsize S2};
% \node[anchor=north, text=red] at (axis cs:3.66, 75.83) {\scriptsize S3};
% \node[anchor=north, text=red] at (axis cs:4.70, 75.54) {\scriptsize S4};

\end{axis}
\end{tikzpicture}
%\caption{Trade-off between F1-score performance and architectural complexity (log-scale parameters). The logarithmic axis highlights that LALE models achieve competitive performance with orders-of-magnitude fewer parameters.}
%\label{fig:f1_vs_params_log}
%\end{figure}
    \end{tabular}
    %\caption{Trade-off between F1-score performance and architectural complexity. Transformer models are given in logarithmic scale due to their large size.}
    \captionof{figure}{Performance (in F1 Score) and architectural complexity comparison. Transformer models are given in log-scale.}
    \label{tab:combined_figures_tradeoff}
\end{table}

\section{Ablation Study}

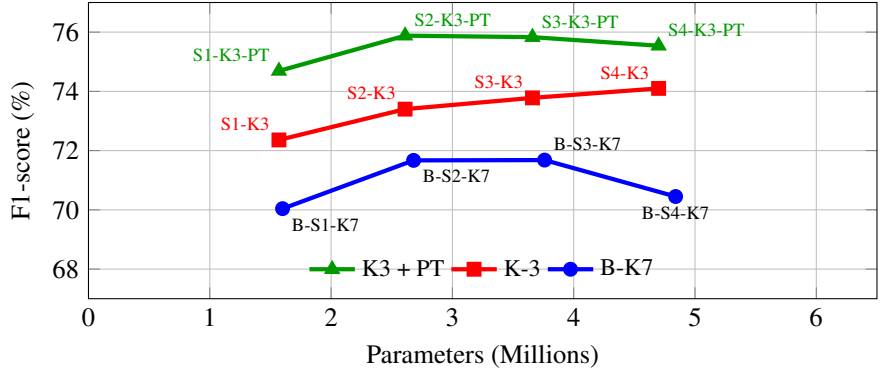
\begin{figure}[ht]
\centering
\begin{tikzpicture}
\begin{axis}[
    width=12cm,
    height=5.5cm,
    xlabel={Parameters (Millions)},
    ylabel={F1-score (\%)},
    grid=major,
    xmin=0, xmax=6.5,
    ymin=67, ymax=77,
    legend style={
        at={(0.5,0.03)},
        anchor=south,
        legend columns=-1,
        nodes={scale=0.9, transform shape},
        draw=none,
        fill=none
    }
]

% --- Node Labels ---
% Proposed (K3 + PT) configurations
\addplot[
    %only marks,
    mark=triangle*,
    color=green!60!black,
    ultra thick,
] coordinates {
    (4.70, 75.54) % S4-K3-PT
    (3.66, 75.83) % S3-K3-PT
    (2.61, 75.88) % S2-K3-PT
    (1.57, 74.69) % S1-K3-PT
};
\addlegendentry{K3 + PT}

% Proposed/S-K3 configurations
\addplot[
    %only marks,
    mark=square*,
    color=red,
    ultra thick,
] coordinates {
    (4.70, 74.10) % S4-K3
    (3.66, 73.78) % S3-K3
    (2.61, 73.40) % S2-K3
    (1.57, 72.36) % S1-K3
};
\addlegendentry{K-3}

% Baseline/B-K7 configurations
\addplot[
    %only marks,
    mark=*,
    color=blue,
    ultra thick,
] coordinates {
    (4.84, 70.45) % B-S4-K7 (BaselineRSFormer-S4)
    (3.76, 71.68) % B-S3-K7 (BaselineRSFormer-S3)
    (2.68, 71.67) % B-S2-K7 (BaselineRSFormer-S2)
    (1.60, 70.04) % B-S1-K7 (BaselineRSFormer-S1)
};
\addlegendentry{B-K7}

% Proposed S-K3-PT labels
\node[anchor=south west, text=green!60!black] at (axis cs:4.70, 75.54) {\scriptsize S4-K3-PT};
\node[anchor=south west, text=green!60!black] at (axis cs:3.66, 75.83) {\scriptsize S3-K3-PT};
\node[anchor=south west, text=green!60!black] at (axis cs:2.61, 75.88) {\scriptsize S2-K3-PT};
\node[anchor=south east, text=green!60!black] at (axis cs:1.57, 74.69) {\scriptsize S1-K3-PT};

% Proposed S-K3 labels
\node[anchor=south east, text=red] at (axis cs:4.70, 74.10) {\scriptsize S4-K3};
\node[anchor=south east, text=red] at (axis cs:3.66, 73.78) {\scriptsize S3-K3};
\node[anchor=south east, text=red] at (axis cs:2.61, 73.40) {\scriptsize S2-K3};
\node[anchor=south east, text=red] at (axis cs:1.57, 72.36) {\scriptsize S1-K3};

% Baseline B-K7 labels
\node[anchor=north] at (axis cs:4.84, 70.45) {\scriptsize B-S4-K7};
\node[anchor=south west] at (axis cs:3.76, 71.68) {\scriptsize B-S3-K7};
\node[anchor=north west] at (axis cs:2.68, 71.67) {\scriptsize B-S2-K7};
\node[anchor=north west] at (axis cs:1.60, 70.04) {\scriptsize B-S1-K7};

\end{axis}
\end{tikzpicture}
\caption{Ablation study showing the performance vs. parameter trade-off. The proposed configurations (S-K3 variants) consistently outperform the baseline configurations (B-K7 variants) across all scales while requiring similar parameters. B = baseline, K\{n\} = kernel size n×n, PT = ImageNet pre-trained, S\{1-4\} = model scale.}\vspace{-10pt}
\label{fig:ablation_f1_vs_params}
\end{figure}

The architecture ablation results in Table~\ref{tab:config_results_ablation} and Figure~\ref{fig:ablation_f1_vs_params} demonstrate the effectiveness of the proposed lightweight architectural improvements. Configurations follow the notation: B = baseline, K\{n\} = kernel size n×n, PT = ImageNet pre-trained, S\{1-4\} = model scale. The optimized LALE models, spanning scales 1 through 4 and utilizing $3\times3$ kernels (S\{1-4\}-K3), consistently achieve higher segmentation performance than the baseline configurations with $7\times7$ kernels (B-S\{1-4\}-K7). By leveraging these smaller kernels alongside efficient design components like StarReLU and RMSNorm, the models simultaneously reduce computational complexity. Furthermore, incorporating ImageNet pre-training for these variants (S\{1-4\}-K3-PT) yields additional performance gains across all scales. Among all configurations, S2-K3-PT achieves the best balance between efficiency and accuracy with only $2.6$M parameters and $0.78$ GMACs.

\begin{table}[ht]
\centering
\small
\renewcommand{\arraystretch}{0.8} 
\caption{Comprehensive performance (\%) and efficiency comparison of different LALE model configurations.}
\label{tab:config_results_ablation}
\begin{tabular}{lccccc|cccc}
\toprule
\textbf{Configuration} 
& \textbf{F1} 
& \textbf{Accuracy} 
& \textbf{Precision} 
& \textbf{Recall} 
& \textbf{IoU} 
& \textbf{Size} 
& \textbf{Throughput} 
& \textbf{Params} 
& \textbf{GMACs} \\

& 
& 
& 
& 
&
& (MB)
& \textbf{(sample/sec)} 
& (M)
& \\
\midrule
S4-K3-PT & 75.54 & 83.82 & 76.20 & 74.93 & 63.27 & 17.94 & 86,537  & 4.7 & 1.16 \\
S3-K3-PT & 75.83 & 84.17 & 76.41 & 75.30 & 63.63 & 13.95 & 104,228 & 3.7 & 0.97 \\
S2-K3-PT & 75.88 & 84.12 & 76.39 & 75.42 & 63.67 & 9.97  & 117,105 & 2.6 & 0.78 \\
S1-K3-PT & 74.69 & 83.16 & 75.18 & 74.26 & 62.25 & 5.99  & 160,253 & 1.6 & 0.59 \\
\midrule
S4-K3    & 74.10 & 82.70 & 74.65 & 73.60 & 61.61 & 17.94 & 75,102  & 4.7 & 1.16 \\
S3-K3    & 73.78 & 82.31 & 73.94 & 73.70 & 61.23 & 13.95 & 115,395 & 3.7 & 0.97 \\
S2-K3    & 73.40 & 82.13 & 73.83 & 73.05 & 60.79 & 9.97  & 118,265 & 2.6 & 0.78 \\
S1-K3    & 72.36 & 81.15 & 72.77 & 72.08 & 59.56 & 5.99  & 164,547 & 1.6 & 0.59 \\
\midrule
B-S4-K7  & 70.45 & 79.76 & 70.98 & 70.09 & 57.46 & 18.49 & 120,117 & 4.8 & 1.48 \\
B-S3-K7  & 71.68 & 80.80 & 72.43 & 71.03 & 58.83 & 14.37 & 134,296 & 3.8 & 1.21 \\
B-S2-K7  & 71.67 & 80.69 & 72.05 & 71.38 & 58.77 & 10.24 & 159,305 & 2.7 & 0.94 \\
B-S1-K7  & 70.04 & 79.15 & 70.62 & 69.63 & 56.93 & 6.12  & 211,432 & 1.6 & 0.67 \\
\bottomrule
\end{tabular}
\end{table}

%% NEW
To assess whether our design generalizes beyond remote sensing, we additionally evaluate LALE on the Liver and Tumor Segmentation Benchmark (LiTS)~\cite{bilic2023liver} dataset and designed two segmentation tasks, liver and tumor segmentation. Our results, available in Table~\ref{tab:combined_segmentation_performance}, are comparable with the reported benchmark for both liver (90\%-96\% F1) and tumor (65\%-74\% F1) segmentation. The experimental results with two different tasks agree with our previous findings and conclusions. On liver segmentation, LALE variants reach within $\sim$1 F1 of the best baseline, consistent with the trade-off observed on ARAS400k. On tumor segmentation, which is substantially more challenging due to severe class imbalance and small lesion size, the gap widens to $\sim$5 F1 against the best baseline. However, LALE remains competitive with many other CNN-based models and the transformer-encoder group and continues to offer the smaller-parameter footprint that motivates the architecture. These results indicate that the resolution-bifurcated hybrid design transfers to medical imaging, though the optimal scale-accuracy operating point appears task-dependent.
%\vspace{-10pt}

\begin{table*}[ht]
\small
\renewcommand{\arraystretch}{0.9}
\setlength{\tabcolsep}{3.5pt} % Adjusts horizontal padding to fit the page
\centering
\caption{Performance comparison of evaluated architectures for liver and tumor segmentation. Performance values in \%.}
\label{tab:combined_segmentation_performance}
\begin{tabular}{lccccccccc}
& \multicolumn{4}{c}{\textbf{Liver Segmentation}} & \phantom{abc} & \multicolumn{4}{c}{\textbf{Tumor Segmentation}} \\
\cmidrule{2-5} \cmidrule{7-10}
\textbf{Architecture} 
& \textbf{F1}  & \textbf{Precision} & \textbf{Recall} & \textbf{IoU} && \textbf{F1} & \textbf{Precision} & \textbf{Recall} & \textbf{IoU} \\
\midrule
DeepLabV3    & 94.68  & 93.91 & 95.46 & 88.52 && 67.02  & 71.39 & 63.15 & 50.40 \\
DeepLabV3+   & 94.82  & 94.03 & 95.62 & 89.45 && 70.24  & 77.44 & 64.26 & 54.13 \\
FPN          & 95.14  & 94.07 & 96.23 & 89.78 && 66.13  & 68.64 & 63.80 & 49.40 \\
Linknet      & 94.93  & 93.64 & 96.25 & 89.21 && 79.47  & 82.18 & 76.94 & 65.94 \\
PAN          & 93.83  & 91.76 & 95.99 & 87.46 && 66.05  & 73.47 & 59.99 & 49.31 \\
Unet         & 95.06  & 93.64 & 96.53 & 90.59 && 79.67  & 82.30 & 77.20 & 66.21 \\
UnetPlusPlus & 95.43  & 94.11 & 96.79 & 90.30 && 73.24  & 83.59 & 65.18 & 57.78 \\
UPerNet      & 94.69  & 93.95 & 95.43 & 89.20 && 71.47  & 79.18 & 65.13 & 55.61 \\
Segformer    & 94.69  & 93.11 & 96.32 & 89.46 && 75.59  & 86.55 & 67.09 & 60.75 \\
\midrule
LALE-S1   & 92.62  & 91.86 & 93.38 & 85.55 && 71.37 & 72.13 & 70.62 & 55.49 \\
LALE-S2   & 93.62  & 93.54 & 93.70 & 88.00 && 69.67 & 65.92 & 73.87 & 53.45 \\
LALE-S3   & 94.13  & 94.09 & 94.17 & 87.92 && 71.71 & 75.27 & 68.47 & 55.90 \\
LALE-S4   & 93.98  & 94.78 & 93.19 & 88.03 && 74.67 & 82.86 & 67.95 & 59.58 \\
\midrule
EffFormer-L3 & 93.40 & 91.63 & 95.24 & 87.52 && 68.05 & 70.67 & 65.61 & 51.57 \\
EffFormer-L7 & 93.93 & 92.09 & 95.83 & 87.82 && 68.86 & 81.76 & 59.47 & 52.51 \\
DeiT3-Base   & 95.13 & 94.30 & 95.97 & 89.40 && 64.91 & 83.06 & 53.27 & 48.05 \\
MaxViT-Tiny  & 94.98 & 93.86 & 96.14 & 90.00 && 60.94 & 66.20 & 56.45 & 43.82 \\
FastViT-SA12 & 92.73 & 91.41 & 94.09 & 86.44 && 68.48 & 73.18 & 64.34 & 52.06 \\
FastViT-MCI0 & 95.01 & 93.86 & 96.19 & 89.90 && 63.17 & 64.62 & 61.78 & 46.16 \\
\bottomrule %\vspace{-25pt}
\end{tabular}
\end{table*}

%% NEW
\section{Training Configuration and Computation Cost}
%Complete-Revised
To create a fair benchmark for all models tried in this work, we utilized the same data augmentation composition, early stopping patience, learning rate decay schedule, loss function (Dice), train-validation-test split, gradient-clipping, efficiency measurements and hardware (same GPU and CPU configuration in the server). We create two benchmarks, one with no pre-trained models, training every model from scratch for ARAS400k dataset, the other with image-net pre-trained models. On a single NVIDIA H100 GPU (80GB) the study with ARAS400k~\cite{ccauglar2026grounding} takes 321 hours in total, 160 hours in architecture search and preliminary trials, 106 hours in baseline and benchmarking and 55 hours for ablation. On average, model training takes 2.2 hours for 80,192 remote sensing images. The extended generalization experiments with LITS~\cite{bilic2023liver} dataset, takes 76 hours in total with 9 hours for liver, 42 hours for tumor segmentation models and 25 hours for architecture search.

\section{Conclusion}
%Complete-Revised
In this work, we introduced LALE, a Lightweight-transformer Architecture for Land-cover Estimation, suitable for training end-to-end remote sensing semantic segmentation models. We designed experiments extensively with ARAS400k, a large-scale remote sensing image segmentation dataset and prepared a benchmark with existing image segmentation models. Our models, specifically tailored to be efficient through integration of convolution layers for high resolution, attention layers for low resolution encoder blocks, a lightweight multi-scale decoder and efficient normalization and activation with RMSNorm and StarReLU respectively. While traditional, parameter-heavy models like UPerNet and Unet retain a marginal lead in the performance, LALE offers comparable performance at the fraction of their size, parameters and operations while providing higher throughput. 

Main advantage of our model is the end-to-end nature designed for remote sensing semantic segmentation task, meanwhile other models require architectural changes, feature extractor encoders or segmentation generation decoders, deteriorating their efficiency with these computation overheads. This work demonstrates that efficient use of attention mechanism and convolution layers, tailored for specific advantages for an optimal approach can create a solid alternative to existing methods. Future work includes deployment-level analysis of LALE for computation constrained hardware, an ablation with different architectural changes to determine most impactful ones and tests with similar datasets.

%\vspace{-10pt}
\acknowledgments
%Complete-Revised
Ümit Mert Çağlar is a beneficiary of the TUBITAK BIDEB 2211-Domestic Graduate Scholarship Program and TUBITAK 2224-A-Grant Program. The experiments reported in this work were fully performed at TUBITAK ULAKBIM, High Performance and Grid Computing Center (TRUBA).
%\vspace{-20pt}
\renewcommand{\refname}{REFERENCES}
\bibliography{main}
\bibliographystyle{spiebib}

\end{document}